\documentclass[onecolumn,aps,pra,reprint,superscriptaddress,longbibliography]{revtex4-1}

\usepackage{amsmath}
\usepackage{graphicx}
\usepackage[lofdepth,lotdepth, caption=false]{subfig}
\usepackage{verbatim}
\usepackage{color}
\usepackage{dcolumn}
\usepackage{bm}
\usepackage{miller}
\usepackage{color}
\usepackage{multirow}

\usepackage{amsfonts}

\usepackage{setspace}

\usepackage{xr-hyper}
\usepackage{hyperref}

\usepackage[english]{babel}

\begin{document}

\title{Layer buckling and absence of superconductivity in LaNiO\textsubscript{2}}

\author{S. Rathnayaka}
\affiliation{University of Virginia, Department of Physics, Charlottesville, Virginia 22904, U.S.A.}

\author{S. Yano}
\affiliation{University of Virginia, Department of Physics, Charlottesville, Virginia 22904, U.S.A.} 
\affiliation{Neutron Group, National Synchrotron Radiation Research Center, Hsinchu, 30076 Taiwan, Republic of China}

\author{K. Kawashima}
\affiliation{Department of Physics and Mathematics, Aoyama Gakuin University, 5-10-1 Fuchinobe, Sagamihara, KANAGAWA 252-5258, Japan}

\author{J. Akimitsu}
\affiliation{Department of Physics and Mathematics, Aoyama Gakuin University, 5-10-1 Fuchinobe, Sagamihara, KANAGAWA 252-5258, Japan}

\author{C.M. Brown}
\affiliation{NIST Center for Neutron Research, National Institute of Standards and Technology, 100 Bureau Drive, MS 6100, Gaithersburg, MD 20899-6100, U.S.A.}
\affiliation{Department of Chemical Engineering, University of Delaware, Newark, DE 19716, U.S.A.}

\author{J. Neuefeind}
\affiliation{Spallation Neutron Source, Oak Ridge National Laboratory, Oak Ridge, TN 37831, U.S.A.}

\author{D. Louca}
\thanks{Corresponding author}
\email{louca@virginia.edu}
\affiliation{University of Virginia, Department of Physics, Charlottesville, Virginia 22904, U.S.A.}

\begin{abstract}
The recent observation of unconventional superconductivity in thin films of LaNiO\textsubscript{2} (critical temperature, T\textsubscript{c}$\sim$10 K) and in bulk single crystals of La\textsubscript{3}Ni\textsubscript{2}O\textsubscript{7} (327) under pressure (T\textsubscript{c}$\sim$80 K), has brought to light a long sought-after class of superconducting nickelates. Through structural measurements in the 327-system, it was shown that the absence of superconductivity is related to bending of the O-Ni-O bonds. Similarly, the bond bending may be linked to the absence of superconductivity in bulk LaNiO\textsubscript{2}. Neutron diffraction was used on bulk non-superconducting La\textsubscript{1-x}Sr\textsubscript{x}NiO\textsubscript{2} samples to show that the layers are naturally buckled, creating a Ni-O-Ni bond angle of 177$^\circ$ at 2 K and ambient pressure. The buckling angle increases to 170$^\circ$ on warming to room temperature. Furthermore, a broad paramagnetic continuum is observed that decreases in intensity on cooling from room temperature signaling a possible transition to a coherent state. However, no antiferromagnetic (AFM) peaks are detected, although enhancement of ferromagnetic (FM) correlations cannot be excluded.  
\end{abstract}

\maketitle

High temperature superconductivity in cuprates whose parent compound is an AFM insulator has driven much of the research in strongly correlated electron systems\cite{bednorz1986possible}. In this class, the CuO\textsubscript{2} plane is a critical component of superconductivity, intimately related to the band position and hybridization between the d\textsubscript{x${^2}$-y${^2}$} and d\textsubscript{z${^2}$-r${^2}$} orbitals that influence the shape of the Fermi surface and the transition temperature. The distance of the Cu ion from the apical oxygen is significant to the hybridization and the transition temperature as predicted by the Maekawa plot\cite{ohta1991apex}. Recently, superconductivity was reported in (La,Sr)NiO\textsubscript{2} films and in Nd doped Nd\textsubscript{1-x}Sr\textsubscript{x}NiO\textsubscript{2} with a T\textsubscript{c}$\sim$10 K\cite{li2019superconductivity,osada2021nickelate}. RENiO\textsubscript{2} (RE – rare earth) with an infinite layered structure (Fig. 1(a)), is isostructural to (Ca,Sr)CuO\textsubscript{2}, the parent compound of Sr\textsubscript{1-x}Nd\textsubscript{x}CuO\textsubscript{2} (T\textsubscript{c}$\sim$34 K) and Sr\textsubscript{1-x}La\textsubscript{x}CuO\textsubscript{2} (T\textsubscript{c}$\sim$43 K). Electronic band structure calculations indicated that the bands near \textit{E}\textsubscript{F} in LaNiO\textsubscript{2} resemble that of CaCuO\textsubscript{2}\cite{sakakibara2010two}. The Ni\textsuperscript{1+} ion in the 3d\textsuperscript{9} electronic configuration is similar to the electronic state of the Cu\textsuperscript{2+} ion with S = 1/2 (see inset of Fig. 1(b) for a simple representation of the orbital states)\cite{siegrist1988parent,azuma1992superconductivity,smith1991electron,er1991superconductivity,lee2004infinite}. Little is known of the contribution of magnetism to the superconducting mechanism while observations of superconductivity in lanthanum doped systems rules out the influence of the RE moment. Resonant inelastic X-ray scattering on NdNiO\textsubscript{3} indicated dispersive excitations with about 200 meV width, reminiscent of spin waves arising from an AFM lattice\cite{lu2021magnetic}.

LaNiO\textsubscript{2} is synthesized by reducing the perovskite LaNiO\textsubscript{3} to the infinite layered structure\cite{crespin1983j,hayward1999sodium,takamatsu2010low}. The electronic configuration of Ni\textsuperscript{1+} (3d\textsuperscript{9}) in the square planar configuration is shown in the inset of Fig. 1(b). According to the Goodenough-Kanamori rules, the interaction of the Ni-O-Ni via 180$^\circ$ degrees ought to be AFM \cite{goodenough1955theory,kanamori1959superexchange} if the spin is localized. A previous neutron diffraction study indicated no long-range magnetic order down to 5 K\cite{hayward1999sodium}. Whether or not a magnetic ground state is present in LaNiO\textsubscript{2} is revisited using both reactor-based and spallation neutron sources. Moreover, the local atomic structure is probed in search for evidence of layer buckling and its implications on Ni-O-Ni coupling. It was previously reported that LaNiO\textsubscript{2} is a Mott insulator in the ground state, and remains paramagnetic down to 5 K\cite{hayward1999sodium,takamatsu2010low}. Our magnetic susceptibility data both for pure and Sr doped LaNiO\textsubscript{2} plotted in Fig. 1(b) show no evidence of a magnetic transition, consistent with earlier measurements\cite{hayward1999sodium}. The Sr-doped samples are also paramagnetic down to low temperatures. 

%Figure 1 for structure
\begin{figure}[t]
\begin{center}
	\centering 
	\includegraphics[width=0.47\textwidth]{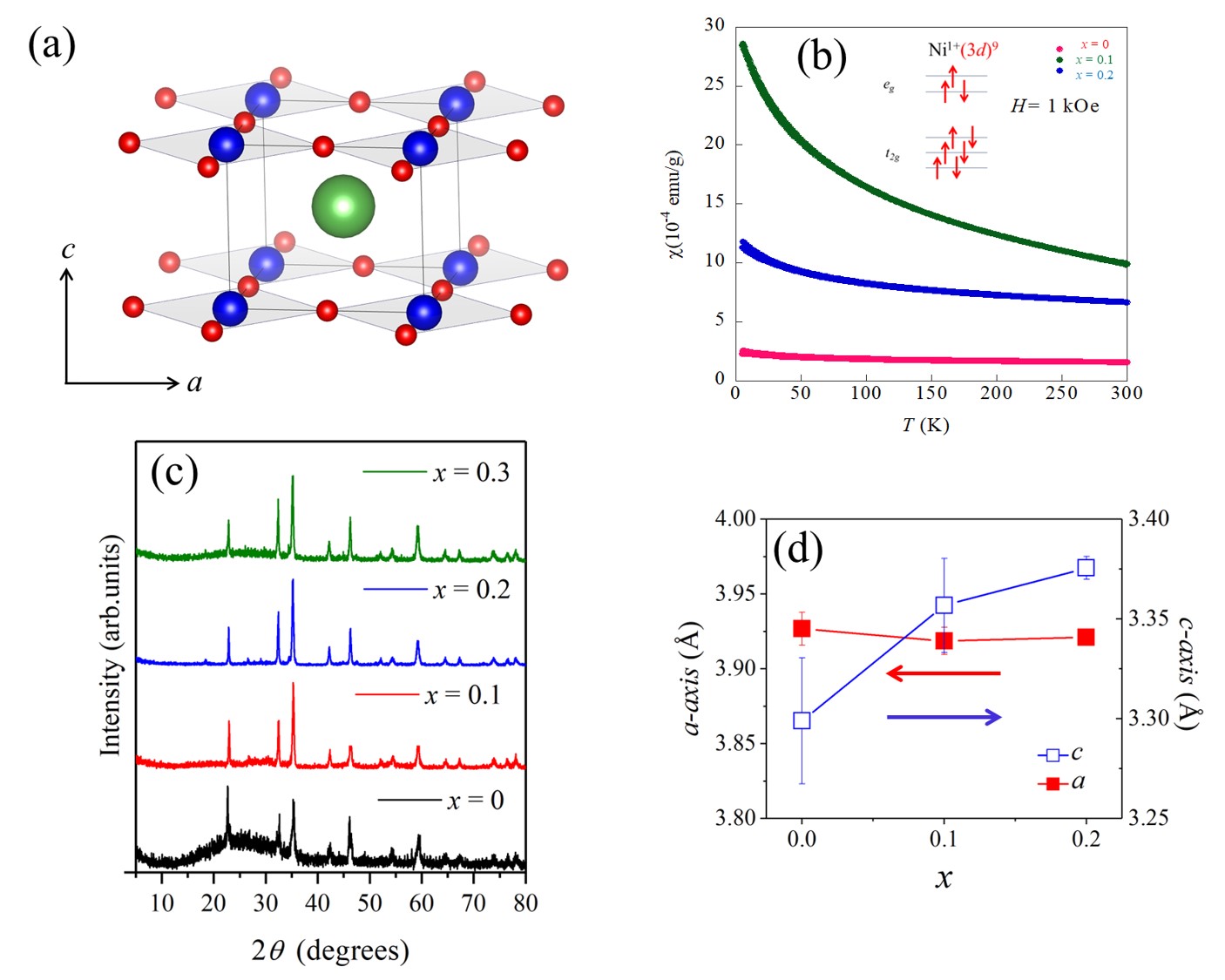}	
 \end{center}
	\caption{(a) The crystal structure of LaNiO\textsubscript{2} with P4/mmm symmetry. (b) The temperature dependence of the magnetic susceptibility of La\textsubscript{1-x}Sr\textsubscript{x}NiO\textsubscript{2} for x = 0, 0.1 and 0.2 under an applied magnetic field of H = 1 kOe (1 Oe = (1000/4$\pi$) A/m, 1 emu/g = 1 Am${^2}$/kg). The inset is a schematic of the electronic configuration of Ni\textsuperscript{1+} in NiO\textsubscript{2} square lattice. (c) Powder X-ray diffraction data of La\textsubscript{1-x}Sr\textsubscript{x}NiO\textsubscript{2} (x = 0, 0.1, 0.2, 0.3). (d) Lattice parameters a and c as a function of Sr\textsuperscript{2+} concentration x. Error bars indicate 1 $\sigma$.} 
 \label{fig1}
\end{figure}

The polycrystalline samples of La\textsubscript{1-x}Sr\textsubscript{x}NiO\textsubscript{2} were synthesized using an alkali molten solution and CaH\textsubscript{2} oxygen reduction method as described in Ref. \cite{crespin1983j}. The sample crystallinity was confirmed using powder X-ray diffraction (XRD), where data were collected on a Multi Flex system with a graphite monochromator. The dc-magnetic susceptibility measurements were performed using a superconducting quantum interference device. The data were collected in field-cooled mode under an applied magnetic field of H = 1 kOe (1 Oe = (1000/4$\pi$) A/m). Specific heat measurements were also performed on the samples in search of phase transitions. 

The neutron scattering experiments were performed at the BT-1 high resolution diffractometer of the NIST Center for Neutron Research (NCNR) \cite{NIST} in the temperature range of 5 to 300 K and at the Nanoscale Ordered Materials Diffractometer (NOMAD) at the Spallation Neutron Source of Oak Ridge National Laboratory. Because of the difficulty of making large amount of power sample ($\sim$2 g), the highest neutron flux condition (monochromator: Ge (311), collimation: 60, wavelength: 2.079 Å) was used at BT-1 while at NOMAD, a white beam was used. The Rietveld refinement\cite{larsongeneral,toby2001expgui} results are summarized in Table I. The NOMAD\cite{neuefeind2012nanoscale} data was normalized, and Fourier transformed to obtain the pair density function (PDF), $\rho(r)$. The PDF is a real-space representation of the atomic correlations. It corresponds to the probability of finding a pair of atoms separated by a distance r by averaging snapshots of pairs over time. Details of this technique can be found in Ref. \cite{toby1992accuracy}. The NOMAD provides high momentum transfer, Q, and data up to 40 Å$^{-1}$ was used for the PDF analysis. The instrumental background and empty can were subtracted from the data, followed by a normalization with vanadium. Corrected data were normalized to obtain the total structure factor, S(Q). The S(Q) was Fourier transformed to obtain the pair correlation function. This method has been applied successfully in many oxide systems\cite{louca1999local,louca1997local}.

\begin{table}[h!]
\caption{The lattice parameters and bond length as a function of temperature. Values in parentheses indicate 1 $\sigma$.}
\centering
\begin{tabular}{c c c c c}
\hline
T(K) & a(Å) & a(Å) & 4*Ni-O(Å) & 8*La-OÅ)  \\ [1ex] % inserts table 

\hline
300 &	3.9544(3)&	3.4158(5)&	1.9772(2)&	2.6133(2) \\
250&	3.9527(3)&	3.4132(5)&	1.9763(1)&	2.6115(2) \\
200&	3.9510(3)&	3.4118(4)&	1.9754(1)&	2.6103(2) \\
150&	3.9497(3)&	3.4071(4)&	1.9749(1)&	2.6084(2) \\
100&	3.9483(3)&	3.4027(5)&	1.9740(2)&	2.6066(3) \\
70&	3.9480(3)&	3.4040(4)&	1.9740(1)&	2.6068(2) \\
50&	3.9477(3)&	3.4023(4)&	1.9739(1)&	2.6060(2) \\
30&	3.9476(3)&	3.4029(3)&	1.9738(1)&	2.6061(2) \\
5&	3.9479(3)&	3.4032(4)&	1.9738(1)&	2.6064(2) \\  [1ex]
\hline
\end{tabular}
\label{table:nonlin}
\end{table}

\begin{table}[h!]
\caption{The calculated momentum transfer, Q, positions and magnetic form factor for potential magnetic Bragg peaks.}
\centering
\begin{tabular}{c c c c}
\hline
(\textit{hkl}) & Q(Å$^{-1}$) & \textit{f}\textsubscript{mag}(Q)\textsuperscript{2} & Reference \\ [0.5ex] % inserts table 

\hline
(0.5 0 0)& 0.7959 & 0.9204 &  \\
(0 0 0.5)& 0.9224 & 0.9077 &  \\
(0.5 0 0.5)& 1.2183 & 0.8266  & SrCuO\textsubscript{2}\cite{goodenough1955theory} \\
\multirow{1}{5.5em}{(0.5 0.5 0.5)} & 1.4553 & 0.7647  & Ca\textsubscript{0.85}Sr\textsubscript{0.15}CuO\textsubscript{2}\cite{vaknin1989antiferromagnetism}, \\ 
&  & & SeFeO\textsubscript{2}\cite{tsujimoto2007infinite}\\
\hline
\end{tabular}
\label{table:nonlin}
\end{table}

Fig. 1(c) is a plot of the X-ray diffraction data for 0 ${<}$ x ${<}$ 0.3 at room temperature. The refinement yields the tetragonal symmetry, P4/mmm\cite{hayward1999sodium}. Pure LaNiO\textsubscript{2} is very close to its presumed stoichiometry, with very little oxygen, if any, at the apical site\cite{hayward1999sodium,alonso1997structural}. Shown in Fig. 1(d) is a plot of the lattice constants as a function of doping. Note that while the a-axis barely changes with doping, the c-axis expands as expected with the Sr substitution. Neutron powder diffraction data collected at the BT-1 powder diffractometer of the NIST Center for Neutron Research (NCNR) on pure LaNiO\textsubscript{2} at 5 and 300 K are shown in Fig. 2. The data at 5 K, collected over a 24-hour period, is compared to the 300 K data, the latter multiplied by a factor of 10 as shown in the intensity versus momentum transfer (Q) plot. The anticipated AFM magnetic peak positions in reciprocal space based on several possible AFM propagation vectors are listed in Table II\cite{zaliznyak1999anisotropic,vaknin1989antiferromagnetism,tsujimoto2007infinite}, considering the magnetic form factor of Ni\textsuperscript{1+} \cite{ref28}. No new Bragg peaks are observed at 5 K that are not otherwise present at 300 K, indicating that no static long-range magnetic order in LaNiO\textsubscript{2} down to 5 K. Magnetic fluctuations may persist however, and this will be discussed below.  

\begin{figure}
% \begin{center}
% 	\includegraphics[width=8.6cm,height=6.5cm]{Fig1.jpg}	
%  \end{center}
	\centering 
	\includegraphics[width=0.47\textwidth]{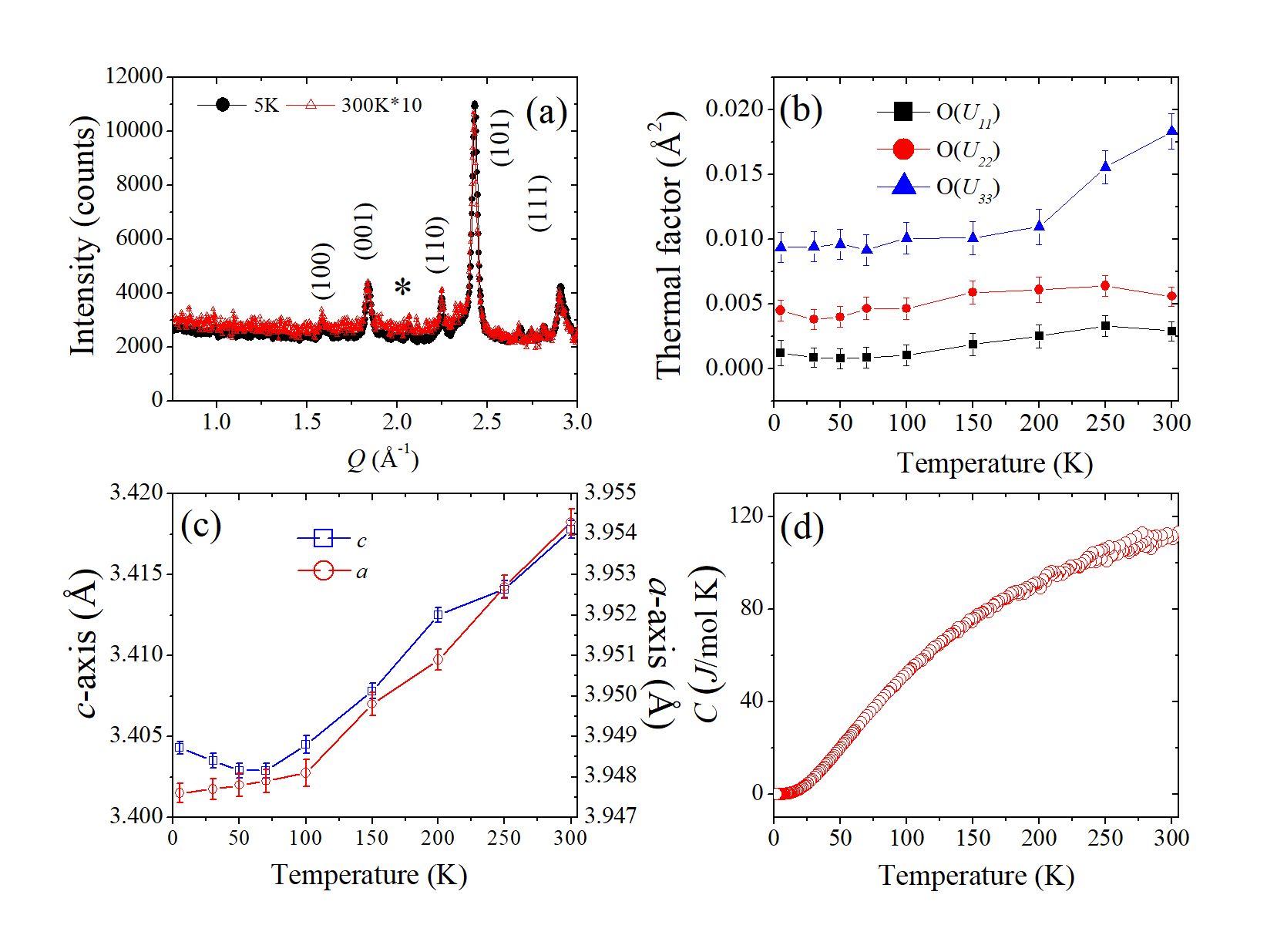}
	\caption{(a) Powder neutron diffraction patterns of LaNiO\textsubscript{2} at 5 K and 300 K from BT-1 (the 300 K data were multiplied by 10). The peaks indexed as * are observed both at 5 K and 300 K. These peaks are close to (0.5 0.5 1) but not exactly. The calculated and observed Q are 2.068 Å$^{-1}$ and 2.155 Å$^{-1}$ at 300 K, 2.070 Å$^{-1}$ and 2.162 Å$^{-1}$ at 5 K. (b) The oxygen thermal factors in LaNiO\textsubscript{2} obtained by the Rietveld refinement method are plotted as a function of temperature. (c) The temperature dependence of the lattice constants a and c for LaNiO\textsubscript{2}. (d) The specific heat of LaNiO\textsubscript{2} as a function of temperature shows no evidence of a transition. Error bars indicate 1 $\sigma$.} 
%	\label{fig_mom0}%
\label{fig2}
\end{figure}

The analysis of the neutron diffraction data also yields the oxygen thermal factors shown in Fig. 2(b). The thermal factor along the z-direction is larger than the ones corresponding to thermal motion in the ab-plane, which may serve as indication of distortions along this direction. The lattice parameters, a and c, linearly increase with increasing temperature in pure LaNiO\textsubscript{2} as shown in Fig. 2(c) where a negative thermal expansion behavior is observed in the c-lattice constant below 50 K. Also shown in Fig. 2(d) is the specific heat that smoothly increases with increasing temperature. With the substitution of Sr\textsuperscript{2+} as in La\textsubscript{1-x}Sr\textsubscript{x}NiO\textsubscript{2} (x = 0.1, 0.2, 0.3), no structural phase transition is observed. The observed diffraction peaks can also be indexed by a tetragonal unit cell (Fig. 1(c)). As seen in Fig. 1(d), the c-lattice parameter increases with the Sr substitution, which is indication that Sr enters the lattice. Attempts to synthesize samples with higher Sr\textsuperscript{2+} concentration were unsuccessful.

%Figure 3 for S(Q)
\begin{figure}[t]
	\centering 
	\includegraphics[width=0.47\textwidth]{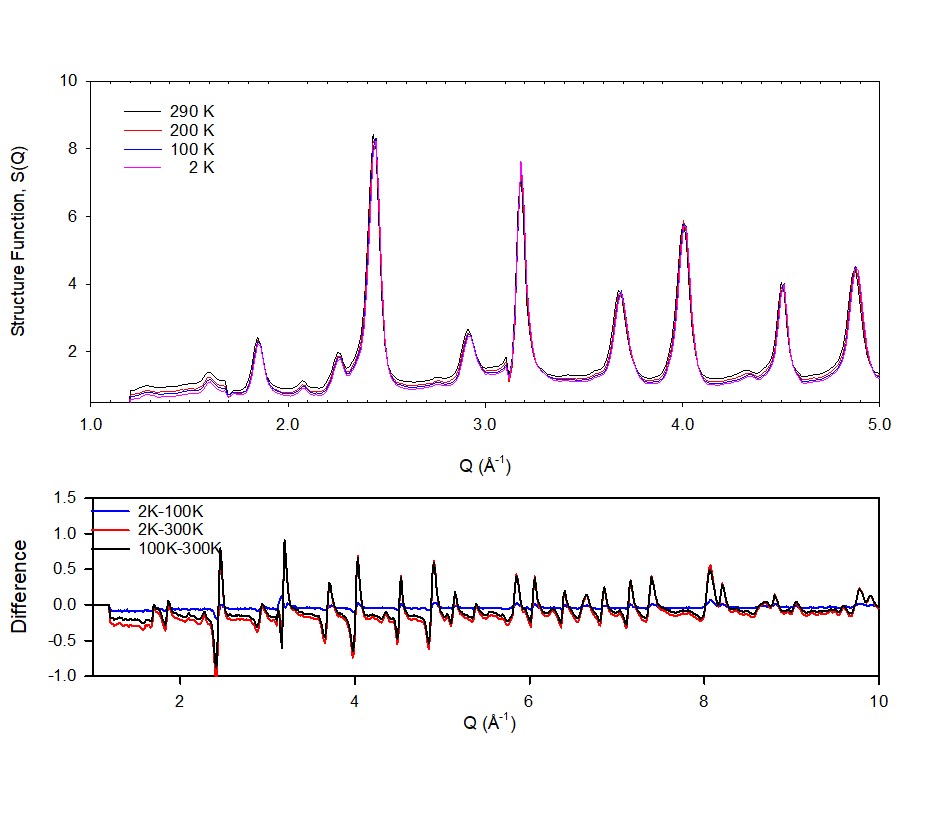}
	\caption{(a) The structure function, S(Q), obtained from the NOMAD neutron diffraction data is plotted at several temperatures. The instrumental background, empty can and normalization with vanadium was performed at all temperatures. The broad diffuse background observed at low Q decreases with cooling. This can be seen in (b) that plots the difference between (2-100)K, (2-300)K and (100-300)K. } 
%	\label{fig_mom0}%
\label{fig3}
\end{figure}

Shown in Fig. 3 is a plot of the structure function, S(Q), obtained from NOMAD, on powder LaNiO\textsubscript{2} as a function of temperature. Even though the instrument background  and empty can were subtracted from the data, the S(Q) exhibits a broad continuum that is temperature dependent. The data shown in the figure are not shifted and the reduction in the background occurs on cooling. The difference in the S(Q) between temperatures is shown in the difference plot below (over a larger Q-range). The changes are more pronounced at low Q. The origin of the broad continuum can be twofold: either it signals the presence of magnetic diffuse scattering or it arises from lattice disorder. In the first scenario, it is reasonable to expect that if the paramagnetic fluctuations are reduced upon cooling, this may lead to an ordered state. No AFM peaks have been detected however i.e. no extra Bragg peaks are evident in the data from both sets of measurements (BT-1 and NOMAD). On the other hand, the condensation of fluctuating spins may produce FM interactions. Given the absence of a magnetic transition in the bulk susceptibility, there is no long-range FM transition, but FM correlations are still possible. Careful examination of the nuclear Bragg peaks at low momentum transfers indicates subtle changes, but it is difficult to discern that any change in the nuclear Bragg intensity is solely due to ferromagnetism. With regards to scenario two, defects or other lattice effects may contribute to the elevated background. While the presence of interstitials and vacancies were ruled out from the refinement of the diffraction data, distortions of the square lattice are possible and discussed next.

%Figure 4 for angle
\begin{figure}[t]
	\centering 
	\includegraphics[width=0.47\textwidth]{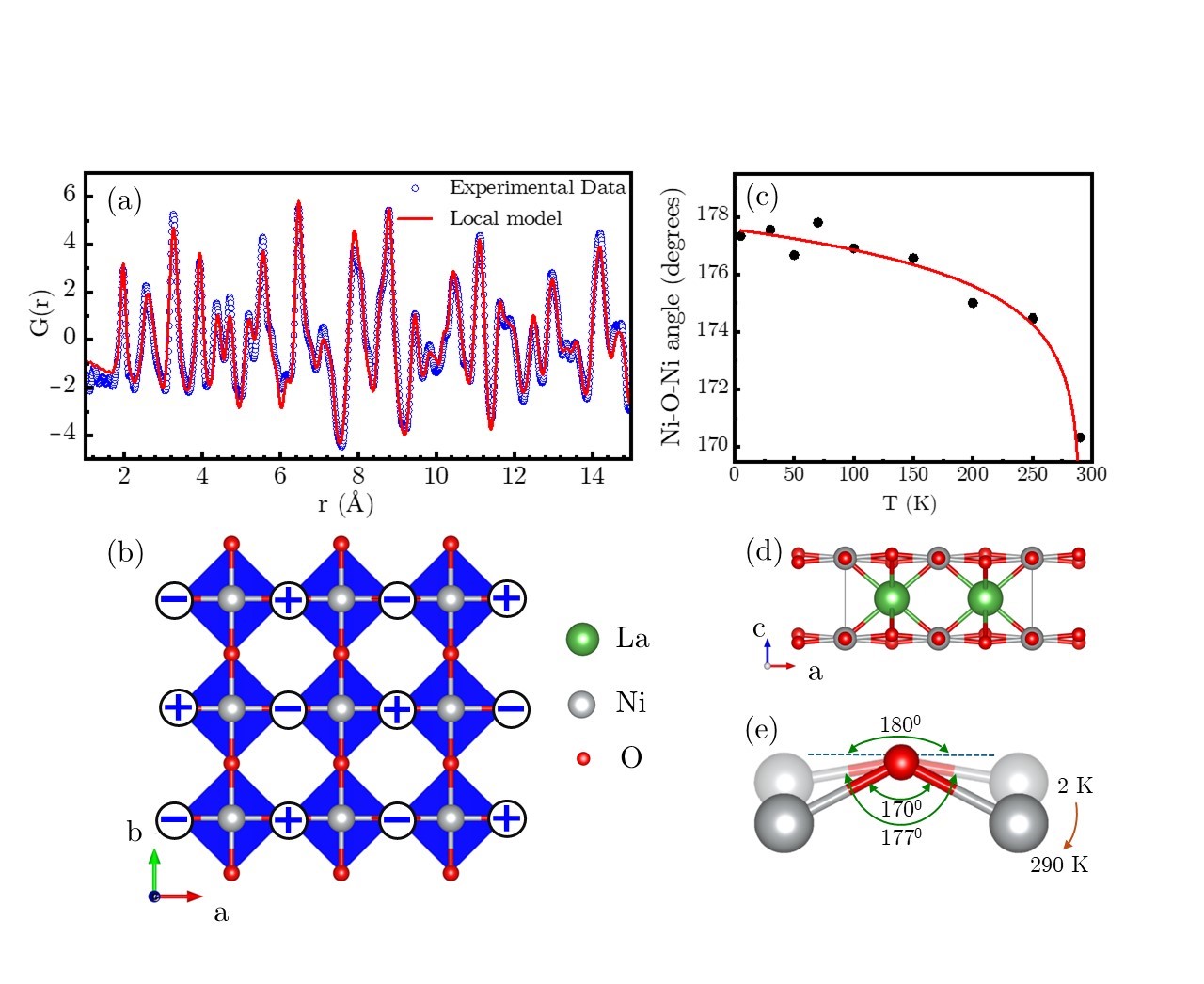}
	\caption{(a) The pair correlation function, G(r), corresponding to the local atomic structure of LaNiO\textsubscript{2} for powder diffraction data collected at 2 K is compared to a local model shown in (b). (b) The local model consists of alternating oxygen vertical displacements along the a-direction. The + and – signs correspond to displacements in the positive or negative z-direction. The displacements are $\approx$ 0.05 Å at 2 K, increasing steadily to 0.17 Å by room temperature. (c) A plot of the Ni-O-Ni bond angle as a function of temperature. Starting at 2 K, the bond is bent away from 180$^\circ$, while the bending continuously increases with warming. (d) and (e) are side views of the bond bending.} 
%	\label{fig_mom0}%
\label{fig3}
\end{figure}

The Fourier transform of the structure function shown in Fig. 3 yields the correlation function, G(r). The G(r) corresponding to the local atomic structure of LaNiO\textsubscript{2} at 2 K is shown in Fig. 4(a). The peaks correspond to the probability of finding a particular atom pair in real-space. The first peak corresponds to the shortest atom-atom correlation length in the crystal and that is from Ni-O pairs. Following are the La-O and O-O correlations and so on. Also shown in the plot is a calculation (solid line) based on a local tetragonal model that shown in Fig. 4(b). In this model, the oxygen atoms are displaced out of plan, in an up/down direction, and consistent with the large z-thermal factors. Specifically, in this model, the oxygen ions along the a- and b-axes are displaced up and down along the c-direction, creating a wave-like motion that buckles the infinite layer in LaNiO\textsubscript{2}. The displacements are antiparallel between sites and are ~ 0.05 Å in magnitude at 2 K. The up/down distortions of oxygen increase to $\approx$0.17 Å by room temperature. This model fits the data very well which indicates that the planes are not flat and buckling of the planes is most likely present (Fig. 4(d/e)). The distortions are analogous to what was observed previously in SrFeO\textsubscript{2}, an AFM insulator, with the same crystal structure (see Refs. \cite{li2016magnetic}). In Fig. 4(c), the temperature dependence of the Ni-O-Ni bond angle is shown. At 2 K, the angle deviates from 180$^\circ$ by 3 degrees. Upon further warming, the buckling increases as shown in Figs. 4(b) and 4(d/e). This behavior was also observed in SrFeO\textsubscript{2}\cite{li2016magnetic}, but the buckling here is less. To conclude, the La\textsubscript{1-x}Sr\textsubscript{x}NiO\textsubscript{2} system was investigated with neutron scattering techniques. The results were complimented by bulk susceptibility and heat capacity measurements. No long-range magnetic order is present although magnetic fluctuations may persist down to low temperatures. Moreover, the infinite layer is buckled as shown by the local structure analysis. This buckling may be contribute significantly to the absence of superconductivity. If superconductivity were to be observed, this system would benefit from a flat NiO\textsubscript{2} plane. We postulate that superconductivity may be observed in thin films as a result of strain induced 180$^\circ$ O-Ni-O bond angles that are conducive to superconductivity. Unbuckling the Ni-O bond was proven to be important to the presence of superconductivity in the 327 system \cite{La3Ni2O7}.

This work has been supported by the Department of Energy, Grant number DE-FG02-01ER45927. We acknowledge the support of the National Institute of Standards and Technology, U. S. Department of Commerce, in providing the neutron research facilities used in this work.  Work at ORNL was supported by the US Department of Energy, Office of Basic Energy Sciences, Materials Sciences and Engineering Division and Scientific User Facilities Division. Support for this work was also provided by the High-tech Research Center Project for Private Universities from the Ministry of Education, Culture, Sports, Science and Technology (MEXT), Grant-in-Aid for Specially promoted Research (JSPS KAKENHI Grant Number 25000003), and Grant-in-Aid for Scientific Research (S) and (B) (JSPS KAKENHI Grant Number 25220803 and 25287087). Moreover, this research was supported by the Strategic International Collaborative Research Program (SICORP (LEMSUPER)) from JST, Japan Science and Technology Agency. Two of authors (KK and JA) were supported by IMRA MATERIAL CO., LTD.

% \section{References}
% \label{Reference}

% \bibliographystyle{plain}
% \bibliography{bibliography.bib}

% \bibliographystyle{elsarticle-num}
\bibliography{bibliography}

\begin{thebibliography}{30}
\bibitem [{NIS()}]{NIST}%
  \BibitemOpen
  \href@noop {} {\bibinfo  {journal} {Certain commercial equipment, instruments, or materials are identified in this document. Such identification does not imply recommendation or endorsement by the National Institute of Standards and Technology, nor does it imply that the products identified are necessarily the best available for the purpose.}\ }\BibitemShut {NoStop}%
\end{thebibliography}

\end{document}